\documentclass[a4paper,12pt]{article}
\usepackage[utf8]{inputenc}
\usepackage{amsmath}
\usepackage{amsfonts}
\usepackage{amssymb}
\usepackage{amsthm}
\usepackage{fontenc}
\usepackage{epstopdf}
\usepackage{epsfig}
\usepackage{tabularx}
\usepackage{graphicx}
\usepackage{array}
\newcolumntype{M}[1]{>{\centering\arraybackslash}m{#1}}
\newtheorem{thm}{Theorem}
\newtheorem{lemma}{Lemma}
\newtheorem{remark}{Remark}
\newtheorem{proposition}{Proposition}

\begin{document}

\title{Radiative gravitational collapse to spherical, toroidal and higher genus black holes}
\author{Filipe C. Mena and Jo\~ao M. Oliveira\\\small{Centro de Matem\'atica, Universidade do Minho,
4710-057 Braga, Portugal}}

\maketitle

\abstract{We derive the matching conditions between FLRW and
generalised Vaidya spacetimes with spherical, planar or hyperbolic
symmetry, across timelike hypersurfaces. We then construct new
models of gravitational collapse of FLRW spacetimes with a negative
cosmological constant having electromagnetic radiation in the
exterior. The final state of the collapse are asymptotically AdS
black holes with spherical, toroidal or higher genus topologies. We
analyse the collapse dynamics including trapped surface formation,
for various examples.
\\\\
Keywords: Black Holes; Gravitational collapse; Spacetime matching;  Exact solutions}

\section{Introduction}
Modeling the process of gravitational collapse to black holes has
been an important challenge in General Relativity and, ever since
the first model of Oppenheimer and Snyder \cite{Opp-Sny}, huge
progress has been made. However, many attempts to build such models
using non-spherical exact solutions of the Einstein field equations
(EFEs) have found no-go results, see e.g. \cite{Mars-Mena-Vera-GRG}
for a review.

The Oppenheimer-Snyder model results from the matching of a
collapsing (spatially homogeneous and isotropic) spherically
symmetric Friedmann-Lema\^itre-Robertson-Walker (FLRW) spacetime to
a (static vacuum) Schwarzschild exterior. As a result of the
matching conditions, the interior must be a dust fluid and,
consequently, the collapse is continuous to a singularity.
Generalizations of this model to spacetimes with a cosmological
constant $\Lambda$ have also, more recently, been constructed
\cite{Nakao}.

The inclusion of the cosmological constant inspired the study of
models of collapse to the so-called {\em topological black holes}
\cite{Smith-Mann, Lemos, Vanzo}. This term has been used to coin
black holes with topologies different from the sphere, e.g. with
toroidal and higher genus topologies. Indeed, a variety of models of
gravitational collapse to topological black holes have been
constructed with $\Lambda$-vacuum exteriors, and interiors given by
FLRW \cite{Smith-Mann, Lemos} and inhomogeneous spacetimes
\cite{MenaCollapse}.

None of the above mentioned models includes radiation in the
exterior. In order to do that using exact solutions of the EFEs, one
can use generalisations of the (radiating) Vaidya metric, such as
the Robinson-Trautman spacetimes, which result from the coupling of
the EFEs to the Maxwell equations in vacuum and can include a
cosmological constant as well as the different spatial topologies
(see e.g. \cite{Kramer}).

For the particular case of Vaidya exteriors (which are spherically
symmetric and have $\Lambda=0$), some interiors are well-known, such
as the FLRW solutions, studied in detail in \cite {FayosVilla}, and
the (inhomogeneous dust) Lema\^itre-Tolman solutions, investigated
in \cite{Gao,Kha}.

Given the importance of the cosmological constant to modern
cosmology and the possibility of the existence of a "landscape" of
vacua states in string theory with $\Lambda$ positive, negative and
zero, it would be interesting to consider radiating exteriors with a
cosmological constant (of any sign) within a class of generalised Vaidya spacetimes, which are sub-classes of the Robinson-Trautman spacetimes.

The global structure of Robinson-Trautman spacetimes with $\Lambda$
has been studied in \cite{Bicak}, where it was observed that these
spacetimes may be used as exact models of black hole formation in
non-spherical settings which are not asymptotically flat. In fact,
the gravitational collapse of Robinson-Trautman spacetimes with
toroidal topology has very recently been analysed in
\cite{VillaVaydia}, where it was shown that incoming electromagnetic
radiation can form black holes as well as naked singularities.
However, as far as we know, the existence of interiors to such
spacetimes has not been studied yet.

In this paper, we intend to construct models of collapse to black
holes by investigating appropriate interiors to sub-classes of
Robinson-Trautman spacetimes. The resulting models, therefore, would
represent the gravitational collapse of an astrophysical object with
a radiating exterior. The black hole would have spherical, toroidal
or higher genus topology depending on the topology of the collapsing
model.

The problem of matching two spacetimes is non-trivial, however, even
for FLRW spacetimes. In fact, it is known that some matchings with
FLRW are impossible: for instance, with a static vacuum
\cite{Seno-Vera,Mars-Mena-Vera-GRG} or with Einstein-Rosen
(gravitational wave) exteriors \cite{Nolan,Tod-Mena}, in cylindrical
symmetry. Indeed, one of the issues has been to ensure that the
matching problem can be solved globally in time, e.g., in some
situations, ensuring that the matching surface maintains its
character during the collapse and that e.g. trapped surfaces, which
appear in the interior, will eventually match exterior ones. Even if
the matching is possible, another problem that may arise is that one
may not be able to construct it from initially untrapped surfaces
(see an example in \cite{Tod-Mena}).

All examples mentioned in the previous paragraph have vacuum
exteriors and, in that case, due to the matching conditions, the
interior radial pressure has to vanish at the matching boundary thus
simplifying the problem. For example, if the interior is FLRW and
the exterior has vacuum then, since the interior has to be dust,
this implies that the boundary is ruled by geodesics which can
considerably simplify the mathematics.

By considering Robinson-Trautman exteriors though, one faces a more
difficult problem since not only the exterior is dynamical, but it
also contains radial pressure. The interior, if FLRW, also needs to
contain radial pressure, which brings extra degrees of freedom into
the problem. Furthermore, while for $\Lambda$-dust matter it is
possible to obtain explicit solutions from Friedmann's equations in
terms of elementary functions, this is not the case, in general, for
FLRW with pressure. In fact, for the hyperbolic FLRW case with
$\Lambda\ne 0$, explicit solutions are only known for few particular
(linear) equations of state, such as radiation fluids
\cite{Griffiths,Harrison,Kramer}. So, the analysis of the matching
problem can also be hampered by that fact.

The plan of the paper is the following: In Section 2, we briefly
revise the spacetimes to be matched, which are the FLRW and the generalised Vaidya spacetimes, both with a cosmological constant and
with the different possible symmetries. In Section 3, we consider
the matching between the two spacetimes, along timelike surfaces of
symmetry, and derive the general matching conditions. In Section 4,
we restrict the problem to $\Lambda<0$, which ensures black hole
formation and specialise the interior to have a linear equation of
state. We then analyse the matched spacetimes for all possible cases
with spherical topology and for some particular cases with toroidal
and higher genus topologies, all satisfying the weak energy
conditions. Finally, Section 5 contains our conclusions.

We use units such that $8\pi G=c=1$, greek indices $\alpha,
\beta,.., \mu,\nu..=0,1,2,3$, latin indices $a,b,c,..i,j,k..=1,2,3$
and capitals $A,B,..=1,2$.
\section{The spacetimes to be matched}
In this section, we briefly review the spacetimes we will attempt to
match.
\subsection{The interior: FLRW with $\Lambda\ne 0$}
For our purposes, it is useful to write the FLRW metric in the form
\begin{equation}
\label{interior} ds^2=-dt^2 + a^2(t)\bigg(dR^2
+f^2(R)\big(dx^2+g^2(x)dy^2\big)\bigg),
\end{equation}
where $a(t)$ is the scale factor and the functions $f(R)$ and $g(x)$
as given in Table \ref{table-FLRW}, for the different values of the
gaussian curvature $k$.
%
%
\begin{center}
\begin{table}[h]
\hspace{1cm}
\begin{tabular}{|l|l|l|}
\hline
$k=1$ & $k=0$ & $k=-1$\\
\hline
$f=\sin R,\cos R;$ & (a) $f=1=g$ & (a) $f=e^{\pm R};\, g=1$\\
$g=\sin x,\cos x$ & (b) $f=R;\, g=\sin x, \cos x$ & (b) $f=\sinh R;\, g=\sin x, \cos x$\\
&&(c) $f=\cosh R;\, g=\sinh x, \cosh x$\\
\hline
\end{tabular}
\caption{The functions $f$ and $g$ of the FLRW metric
(\ref{interior}) depending on the curvature $k$.} \label{table-FLRW}
\end{table}
\end{center}
All cases with $k=1$ correspond to spherical symmetry. For $k=0$
or $-1$, cases (a) have planar symmetry, (b) spherical symmetry and
(c) hyperbolic symmetry.
From the EFEs, we recall that one gets
\begin{equation}
\label{density} \rho = \frac{3(a_{,t}^2 + k)}{a^2}-\Lambda,
~~~~~~~~p = -\frac{2a_{,{tt}}}{a}-\frac{a_{,t}^2 + k}{a^2}
+\Lambda,\end{equation} which are the Friedmann equations.

It will be important below to check when the 2-surfaces of constant
$t$ and $R$ are trapped or marginally trapped. We then take two null
normals $\vec k^\pm$ to these 2-surfaces satisfying
$\vec{k}^\pm\cdot\vec{k}^\pm = 0,~~\vec{k}^\pm\cdot \vec{e}_A
=0,~~\vec{k}^+\cdot \vec{k}^- = -1$, where $\vec{e}_A$ are tangent
vectors to the surfaces, and calculate the null expansions $\theta^{\pm}
= \eta^{AB}\theta^{\pm}_{AB} = -\eta^{AB}k_\mu^{\pm}e_A^\nu\nabla_\nu
e_B^\mu$, where $\eta^{AB}$ is the respective 2-metric, giving
$\theta^\pm = (\sqrt{2}/af)(a_{,t}f\pm f')$ and
\begin{equation}
 \theta^+\theta^- = \frac{2}{a^2f^2}(a_{,t}^2f^2-(f')^2),
\end{equation}
where the prime denotes differentiation with respect to $R$.  As
standard, we say that the 2-surfaces are trapped if
$\theta^+\theta^->0$ and marginally trapped when
$\theta^+\theta^-=0$. We call {\em apparent horizon} to a future
marginally trapped surface.

We also recall that the {\em weak energy conditions}, in this case, are
$\rho\ge 0$ and $\rho+p\ge 0$, while the {\em dominant energy
conditions} are the weak conditions together with $\rho\ge p$. Later
on, we will also assume a linear equation of state of the form $p =
\gamma \rho$, where $\gamma$ is a constant, in which case
Friedmann's equations \eqref{density} reduce to:
\begin{equation}\label{eqa}
  \frac{a_{,tt}}{a} = -\frac{a_{,t}^2 + k}{2a^2}(1+3\gamma) +
  \frac{\Lambda}{2}(1+\gamma)
\end{equation}
and the dominant energy conditions imply $1\ge\gamma\ge -1.$
\subsection{The exterior: Generalised Vaidya spacetimes}\label{secext}
We consider spacetime metrics in the form:
\begin{equation}
\label{exterior} ds^2= -\chi du^2 + 2\varepsilon dudr +
r^2\big(d\theta^2 + \Sigma^2(\theta)d\phi^2\big),
\end{equation}
where
$$\chi= b - \frac{2m(u)}{r} - \frac{\Lambda }{3}r^2$$
and $u$ is a null coordinate, which for $\varepsilon = +1$ is
advanced and for $\varepsilon=-1$ is retarded. In turn, $m(u)$ is
the mass function and $\Sigma(\theta)=\sin\theta$, $\sinh\theta,
\theta$ for $b=1,-1,0,$ i.e. for spherical, hyperbolic and planar
geometry, respectively. The particular case with $b=1$ and
$\Lambda=0$ corresponds to the well-known Vaidya solution.

The above metric admits a geodesic, shear-free, twist-free but
expanding null congruence \cite{Griffiths}. It corresponds, in
general, to aligned pure radiation (i.e. a flow of matter of zero
rest-mass propagating in the principal null direction) with
$T_{\alpha\beta}=\rho(u,r) k_\alpha k_\beta$, where ${\bf k}=-du$ is
a null one form. A detailed physical interpretation of this metric
as a solution of the Einstein-Maxwell equations with $\Lambda$ for a
null electromagnetic field is given in \cite{Kramer, VillaVaydia}.

The spacetime has three Killing vectors, in general, and the
corresponding isometry group acts transitively on the spacelike
2-surfaces with constant $u$ and $r$. These surfaces of transitivity
can describe spheres for $b=1$, tori for $b=0$ or higher genus
surfaces for $b=-1$, with the appropriate identifications. The
dominant energy condition, in this case, gives (see also
\cite{FayosVilla}):
\begin{equation}\label{wec}
 \varepsilon \frac{dm}{du}\geq 0.
\end{equation}
It is important to note that for incoming (outgoing) radiation, we
can only have increasing (decreasing) $m$ for an increasing $u$. As
we would like to think of $m$ as increasing with incoming radiation
($\varepsilon=1$) and decreasing with outgoing radiation
($\varepsilon=-1$), we will consider that $u$ grows towards the
future (as in \cite{FayosVilla}).

The null expansions of the 2-surfaces of transitivity, in this case,
are $\theta^-= -2\varepsilon/r$ and $\theta^+=\varepsilon \chi/r$,
so that
$$
\theta^+\theta^-=-\frac{2\chi}{r^2}
$$
and those 2-surfaces are trapped as long as $\chi<0$. The surfaces
are future trapped or past trapped for $\varepsilon=1$ or
$\varepsilon=-1$, respectively.

In turn, for $\chi=0$ and $\varepsilon=1$, there are apparent
horizons (AH)\footnote{In fact, these are {\em marginally trapped
tubes} or {\em apparent 3-horizons} \cite{TSurfaces}, which are
hypersurfaces foliated by future marginally trapped surfaces. To
simplify the terminology, we simply call them apparent horizons.}
given by:
\begin{equation}
 \frac{\Lambda r_H^3}{3} - br_H + 2m(u) = 0
\end{equation}
and their causal character is determined by the sign of
\begin{equation}
\varepsilon\frac{dr}{du}= \frac{2\varepsilon}{b-\Lambda
r^2}\frac{dm}{du}.
\end{equation}
So, the existence of apparent horizons as well as its causal
character are constrained in the three cases $b=0,-1,1$ as follows
(assuming the dominant energy condition holds):
\begin{itemize}
\item For $b=1$, assuming $dm/du\neq0$, we just need to look at the sign of
$1-\Lambda r^2$. For $\Lambda\leq0$, the AH is spacelike, while for
$\Lambda>0$ it is spacelike for $1>\Lambda r^2$ and timelike for
$1<\Lambda r^2$. When the AH is spacelike it is called
\emph{dynamical horizon} \cite{Ashtekar, VillaVaydia, TSurfaces}. If
$dm/du=0$, the AH is a null surface.

\item For $b=0$, the existence of solutions to $\chi=0$ implies that
either $m$ or $\Lambda$ is negative. With $dm/du=0$, we again have a
null AH, while for $dm/du\neq0$ the AH is timelike for $\Lambda>0$
and spacelike for $\Lambda<0$. If $\Lambda=0$, then $\chi=0$ implies
$m(u)=0$ which corresponds to a null AH.

\item For $b=-1$, we must have $m$ or $\Lambda$ negative.
For $dm/du\neq0$ and $\Lambda\geq0$, the AH is always timelike,
while for $\Lambda\leq0$ it is timelike for $1>|\Lambda|r^2$ and
spacelike for $1<|\Lambda|r^2$. Just like in the other two cases,
the AH is null for $dm/du=0$.
\end{itemize}
The causal character of the AH is important since it will help to
explain the dynamics studied in Section 4.

Finally, we recall that for a constant $m$, the metric \eqref{exterior} reduces to the Kottler (vacuum static) metric
\begin{equation}
ds^{2}=-\chi dT^2+\chi
^{-1}dr^2+r^2(d\theta^2+\Sigma^2(\theta)d\varphi^2). \label{K1}
\end{equation}
In this case, for $\Lambda<0$ and $m>0$, $\chi$ has a unique
positive zero and this solution describes a black hole with planar,
spherical or hyperbolic symmetry, respectively, on an asymptotically
AdS background. For $b=0$ or $-1$ it is possible to make
identifications of the 2-metric of constant $T$ and $r$ to obtain
toroidal and higher genus black holes.
\section{The matching conditions}\label{secm}
Given spacetimes $(M^\pm, g^\pm)$ with non-null boundaries
$\sigma^\pm$, matching them requires an identification of the
boundaries, i.e. a pair of embeddings $\Phi^\pm: \sigma\to M$ with
$\Phi^\pm(\sigma)=\sigma^\pm$, where $\sigma$ is an abstract copy of
either boundary. Let $\xi^i$ be a coordinate system on $\sigma$.
Tangent vectors to $\sigma^\pm$ are obtained by $\vec
e_i=\partial\Phi/\partial\xi^i$. There exist also unique (up to
orientation) unit normal vectors $n^\alpha_\pm$ to the boundaries.
The first and second fundamental forms of $g^\pm$ on $\sigma^\pm$
are given by
$$
\gamma^\pm_{ij}=e^{\pm\alpha}_i e^{\pm \beta}_j
g_{\alpha\beta}|_{\sigma^\pm},~~~~ K^\pm_{ij}=-n^{\pm}_{\alpha}
e^{\pm\beta}_i \nabla^{\pm}_\beta e^{\pm\alpha}_j.
$$
The matching conditions (in the absence of shells) require the
equality of the first and second fundamental forms on $\sigma^\pm$,
i.e.
$$
\gamma^+_{ij}=\gamma^-_{ij},~~~~ K^+_{ij}=K^-_{ij}.
$$
If spacetime symmetries are present, one chooses the $\vec e^{~{\pm}}_i$
to reflect the symmetries so that the expressions for $K^\pm_{ij}$
simplify.

Our objective is to match metric (\ref{exterior}) as an exterior to
a FLRW interior (\ref{interior}). The matching will be considered
along timelike surfaces with coordinates $\xi^i
=\{\tau,\vartheta,\varphi\}$ parametrized as
\begin{eqnarray}
\label{parametrization}
\sigma^+&=&\{u=u(\tau),~ r=r(\tau),~ \theta=\vartheta\,~ \phi=\varphi\} \nonumber\\
\sigma^-&=& \{t=t(\tau),~ R=R(\tau),~ x=\vartheta,~ y=\varphi\}
\end{eqnarray}
The vectors fields $\vec{e}_i^\pm$, generators of the surfaces
$\sigma^\pm$, can  be written as
\begin{equation}
\begin{split}
 &\vec{e}^{+}_\tau = \dot{u}\partial_u + \dot{r}\partial_r \\
 &\vec{e}^+_\vartheta = \partial_\theta \\
 &\vec{e}^+_\varphi = \partial_\phi \\
 \end{split}
 \hspace{2cm}
 \begin{split}
 &\vec{e}^-_\tau = \dot{t}\partial_t + \dot{R}\partial_R\\
 &\vec{e}^-_\vartheta = \partial_x,\\
 &\vec{e}^-_\varphi = \partial_y,\\
 \end{split}
\end{equation}
where a dot denotes derivatives with respect to $\tau$. In turn, the
normal vectors to $\sigma^\pm$, which satisfy $n_\mu^\pm
e_a^{\pm\mu}=0$ and $n^\pm_\mu n^{\pm\mu}=1$, give
\begin{equation}
\begin{split}
\label{normals}
 &\vec{n}^+ = \frac{\overline{\epsilon}}{\sqrt{\chi \dot{u}^2-2\varepsilon\dot{u}\dot{r}}}
 \bigg(\varepsilon \dot{u} \partial_u + (\chi\dot{u}-\varepsilon\dot{r})\partial_r\bigg),\\
 &\vec{n}^- =
 \frac{\epsilon}{\sqrt{\dot{t}^2-a^2\dot{R}^2}}\bigg(a\dot{R}\partial_t+\frac{\dot{t}}{a}\partial_R\bigg),
 \end{split}
\end{equation}
where the constants $\overline{\epsilon}$ and $\epsilon$ satisfy
$\overline{\epsilon}\epsilon = \pm 1$, which comes from the fact that
the normal vectors are defined up to a sign\footnote{Here, we keep
the notation of \cite{FayosVilla} and the constant $\epsilon$
shouldn't be confused with $\varepsilon$.}. With this framework, we
can now obtain the first fundamental forms on $\sigma^\pm$ as
\begin{equation}\label{fform}
\begin{split}
 &\gamma^+ = (-\chi\dot{u}^2 + 2\varepsilon\dot{u}\dot{r})d\tau + r^2\big(d\vartheta^2 +\Sigma^2(\vartheta) d\varphi^2\big),\\
 &\gamma^- = (-\dot{t}^2+a^2\dot{R}^2)d\tau^2 + a^2f^2\big(d\vartheta^2 + g^2(\vartheta)d\varphi^2\big).
\end{split}
\end{equation}
The equality of the first fundamental forms thus gives $g(\vartheta) =
\Sigma(\vartheta)$ together with:
\begin{equation}\label{gm}
\begin{cases}
 \chi\dot{u}^2 - 2\varepsilon\dot{u}\dot{r} \stackrel{\sigma}{=} \dot{t}^2-a^2\dot{R}^2\\
 r^2 \stackrel{\sigma}{=} a^2f^2.
\end{cases}
\end{equation}
where $\stackrel{\sigma}{=}$ stands for an equality on the
surface $\sigma$.
\begin{remark}
In most cases we omit the symbol $\stackrel{\sigma}{=}$ as it should
be clear from the context whether a relation (equality or
inequality) is applied on $\sigma$. For example, whenever a dot
derivative is involved, then the relation is applied on $\sigma$
since that derivative is only defined on $\sigma$.
\end{remark}
We can now calculate the second fundamental forms on $\sigma^\pm$
with:
\begin{equation}
 K^{\pm}_{ij} = -n_{\pm\mu}\bigg(\frac{\partial^2 \Phi^\mu}{\partial\xi^i\partial\xi^j}
 + \Gamma_{\pm\gamma\lambda}^\mu\frac{\partial \Phi^\gamma}{\partial \xi^i}\frac{\partial \Phi^\lambda}{\partial \xi^j}\bigg),
\end{equation}
giving the non-zero components
\begin{equation}
 \begin{split}
  &K^+_{\tau\tau} = \frac{\overline{\epsilon}}{\sqrt{\chi \dot{u}^2-2\varepsilon\dot{u}\dot{r}}}
  \bigg[\dot{r}\bigg(\ddot{u}+\frac{\varepsilon}{2}\dot{u}^2\chi_{,r}\bigg)
  -\dot{u}\bigg(\ddot{r}-\frac{\varepsilon}{2}\dot{u}^2\chi_{,u} + \frac{1}{2}\dot{u}^2\chi\chi_{,r}
  -\varepsilon\dot{u}\dot{r}\chi_{,r}\bigg) \bigg]\\
  &K^+_{\vartheta\vartheta} = \frac{1}{\Sigma^2}K^+_{\varphi\varphi} = \frac{\overline{\epsilon}}{\sqrt{\chi \dot{u}^2-2\varepsilon\dot{u}\dot{r}}}
  (\chi\dot{u}-\varepsilon\dot{r})r,\\
  &K^-_{\tau\tau} = \frac{\epsilon}{\sqrt{\dot{t}^2-a^2\dot{R}^2}}\bigg[a\dot{R}\bigg(\ddot{t}+a_{,t}a\dot{R}^2\bigg)
  -a\dot{t}\bigg(\ddot{R}+2\frac{a_{,t}}{a}\dot{R}\dot{t}\bigg)\bigg],\\
  &K^-_{\vartheta\vartheta} = \frac{1}{g^2}K^-_{\varphi\varphi} = \frac{\epsilon}{\sqrt{\dot{t}^2-a^2\dot{R}^2}}
  \bigg(a^2\dot{R}a_{,t}f^2 + \dot{t}aff'\bigg).\\
 \end{split}
\end{equation}
Matching the second fundamental forms gives us:
\begin{equation}\label{Km1}
  \dot{r}\ddot{u}-\dot{u}\ddot{r} + \frac{\varepsilon}{2}\dot{u}^2\dot{r}\chi_{,r}+\frac{\varepsilon}{2}\dot{u}^3\chi_{,u}
  -\frac{1}{2}\dot{u}^3\chi\chi_{,r} + \varepsilon \dot{u}^2\dot{r}\chi_{,r} \stackrel{\sigma}{=} \overline{\epsilon}\epsilon\big(
  a\dot{R}\ddot{t} + a^2a_{,t}\dot{R}^3 - a\dot{t}\ddot{R} - 2a_{,t}\dot{R}\dot{t}^2\big),
\end{equation}
\begin{equation}\label{Km2}
  \chi\dot{u}-\varepsilon\dot{r} \stackrel{\sigma}{=} \overline{\epsilon}\epsilon\big(aa_{,t}\dot{R}f+\dot{t}f'\big) .
\end{equation}
After quite long calculations to simplify equations \eqref{gm},
\eqref{Km1} and \eqref{Km2}, we end up with:
\begin{eqnarray}\label{equ}
 \dot{u} &\stackrel{\sigma}{=}& \varepsilon\frac{\dot{t} + \varepsilon\overline{\epsilon}\epsilon a\dot{R}}
 {\varepsilon\overline{\epsilon}\epsilon f' - a_{,t}f},\\
\label{eqr}
 r &\stackrel{\sigma}{=}& af,\\
\label{eqm}
 m &\stackrel{\sigma}{=}& \frac{1}{2}aa_{,t}^2f^3 - \frac{1}{2}aff'^2 +
 \frac{b}{2}af-\frac{\Lambda}{6}a^3f^3,\\
\label{eqR}
 \dot{R} &\stackrel{\sigma}{=}&
 \varepsilon\overline{\epsilon}\epsilon\frac{p}{a\rho}\dot{t},
\end{eqnarray}
where the interior pressure $p$ and density $\rho$ are defined by
the Friedmann equations \eqref{density}. We note that for $b=1$ and
$\Lambda=0$, equations \eqref{equ}-\eqref{eqR} reduce to the ones of
\cite{Fayos-primeiro-artigo}.

The requirement for the matching hypersurface to be timelike
implies, using the first fundamental forms \eqref{fform}, the
inequalities
\begin{equation}
\label{ineq}
 \begin{split}
  &\chi\dot{u}^2-2\varepsilon\dot{r} \dot{u}>0,\\
  &\dot{t}^2 - a^2\dot{R}^2 > 0.\\
 \end{split}
\end{equation}
The above analysis leads to the following result:
\begin{thm}
The necessary and sufficient conditions to match the spacetimes
(\ref{interior}) and (\ref{exterior}), across a timelike
hypersurfaces parametrized by \eqref{parametrization}, are given by
equations \eqref{equ}-\eqref{eqR} together with the inequalities
\eqref{ineq}.
\end{thm}
\begin{remark}
We note that the matching hypersurface should be determined by
solving the matching conditions in each case. This is one of the goals
of Section 4, where we consider more specific forms for the metrics.
\end{remark}
In our matching procedure, we match the same kind of spatial
topology in the interior and exterior spacetimes, so that
\begin{equation}
\label{useful}
f'^2=b-kf^2
\end{equation}
is satisfied for the possible combinations of Table 1, implying $f''=-kf$ and,
then, any derivative of $f$ can be written in terms of the $f$
functions themselves.
\begin{remark}
From \eqref{eqm} and \eqref{useful}, we get
\begin{equation}
\label{masscontinuity} m \stackrel{\sigma}{=}{\displaystyle
\frac{a^3f^3\rho}{6}},
\end{equation}
which makes transparent an important physical consequence of the
matching conditions, which is the continuity of the mass across
$\sigma$.
\end{remark}
\subsection{Case $p=\gamma\rho$}
Assuming a linear equation of state $p=\gamma\rho$, the inequalities
\eqref{ineq} give
\begin{equation}
\label{timelike}
 \begin{split}
  &1>\frac{p^2}{\rho^2} \Rightarrow 1 > \gamma > -1,\\
 \end{split}
\end{equation}
which, for $\dot u>0$, implies from \eqref{equ}
\begin{equation}\label{spcond}
\overline{\epsilon}\epsilon f' > \varepsilon a_{,t}f,
\end{equation}
restricting the $\gamma$ values for each functions $f$ and $a$. Note
that \eqref{timelike}, together with $\rho>0$, ensures the dominant
energy conditions.

It is also easy to check that
\begin{equation}\label{mconst}
 m \stackrel{\sigma}{=} const \iff \gamma=0,
\end{equation}
and, when $m$ is constant in the whole exterior spacetime (not just
on the surface $\sigma$), we recover the results of
\cite{MenaCollapse} for the matching of a dust interior and a
Kottler exterior. To prove \eqref{mconst}, we need to obtain $\rho$
as a function of $a$. With the Friedmann equations \eqref{density},
it is well known that we can obtain such an equation as $\rho
a^{3(1+\gamma)} = \rho_0,$ where $\rho_0> 0$ is a constant. So, for
$\gamma=0$, we have $m \stackrel{\sigma}{=} f^3 \rho_0/6$ and, from
\eqref{eqR}, $\dot R\stackrel{\sigma}{=} 0$, which means that $f(R)$
is also constant and we recover result \eqref{mconst}.
\subsection{Cases $\Lambda\ge 0$}
Before proceeding, some comments are in order regarding the cases $\Lambda=0$ and $\Lambda>0$:
\begin{itemize}
\item [{\bf (i)}] {\bf Case ${\bf \Lambda=0}$}:
Considering $m>0$, the cases $b=0$ and $b=-1$ can be excluded, {\em
a priori}, since they wouldn't represent black hole formation as
explained in Section 2.2.
In turn, the case $k=0$ and $b=1$ can be matched and was analysed in
great detail in \cite{FayosVilla}.

\item [{\bf (ii)}] {\bf Case ${\bf \Lambda>0}$}: In this case, the gravitational collapse results, in
most cases, on a bounce that prevents the singularity formation. So,
we will not treat this case here and leave it for a future
publication.
\end{itemize}
We shall then proceed with the case $\Lambda<0$ which, as we shall
see, can represent the gravitational collapse of a FLRW fluid to an
asymptotically AdS black hole with either spherical, toroidal or
higher genus topology.
\section{Analysis of the matched spacetimes for $\Lambda<0$}
To start with, we observe that if we assume that the initial
hypersurface is untrapped, then the matching for the case $k=b=0$
and $\Lambda< 0$ is impossible since, from the matching conditions,
we get
\begin{equation}
\label{chieq} \chi \stackrel{\sigma}{=} (f')^2-\frac{\dot a^2
f^2}{\dot t^2}=-\frac{\dot a^2}{\dot t^2}<0.
\end{equation}
We highlight this in the following lemma:
\begin{lemma}
It is impossible to match metrics (\ref{interior}) and (\ref{exterior}) with $m>0$, initially untrapped, across a timelike
hypersurface, as in Theorem 1, for $k=b=0$ and $\Lambda< 0$.
\end{lemma}
There is a similar result for a FLRW interior and a Kottler exterior
\cite{Smith-Mann}, so this result might not be surprising.

We shall then consider other values of $k$ and $b$, giving examples
of spacetime matchings between FLRW interiors and 
the generalised Vaidya exteriors given by \eqref{exterior} for the three different topologies. In order to do that,
we shall assume, from now on, the linear equation of state $p=\gamma
\rho$ and, without loss of generality, $t(\tau)=\tau$ at the
matching boundary. The same choice was used by the authors of
\cite{FayosVilla} and we intend to make a close comparison of our
results with theirs, at least in the spherical case.
For that case, we summarize as follows the main results which are
proved in Section 4.1.:
\begin{thm}
Given a spherically symmetric interior FLRW spacetime $(M^-,g^-)$
with $k=0$, $0<\gamma<1$ and $\Lambda<0$, there is always a
one-parameter family of timelike surfaces of symmetry $\sigma$,
across which it is possible to match $(M^-,g^-)$ to a
spacetime $(M^+,g^+)$ having the metric \eqref{exterior} with $b=1$, for a time
interval depending on $\varepsilon$ and $\overline \epsilon
\epsilon$. The matched spacetime satisfies the dominant energy
conditions.

Given $(M^-,g^-)$ and the free parameter at $\sigma$, $(M^+,g^+)$ is
uniquely determined by the matching conditions. For $\varepsilon=1$
and $\overline \epsilon \epsilon=1$, there are open sets of initial
data such that the matched spacetimes are initially untrapped and
eventually (re)collapse forming an apparent horizon and an
asymptotically AdS black hole with spherical topology.
\end{thm}
While in the spherical case with $k=0$, one can write the solutions
to the Friedmann equation in terms of elementary functions for any
$\gamma$, this is not the case for $k=-1$ and $\Lambda<0$ where,
instead, particular $\gamma$ cases will be analysed. Such particular
solutions, satisfying the dominant energy conditions, can be
obtained for $\gamma=-2/3,-1/3,1/3$. A summary of the main results
of sections 4.2 and 4.3 is as follows:
\begin{proposition}
It is possible to match an interior FLRW spacetime \eqref{interior} with
$k=-1$ and $\Lambda<0$, across timelike surfaces of symmetry
$\sigma$, to an exterior spacetime \eqref{exterior}, satisfying
the weak energy conditions, for:
\begin{itemize}
\item $b=0$, toroidal topology and
$\gamma=1/3$ or $\gamma=-2/3$, if $\varepsilon=1,\overline \epsilon
\epsilon= 1$ or $\varepsilon=-1, \overline \epsilon \epsilon=1$,
respectively.
\item $b=-1$, higher genus topology and $\gamma=1/3$, if $\varepsilon=1,\overline \epsilon
\epsilon=1$.
\end{itemize}
In the cases with $\varepsilon=1$, there are open sets of initial
data such that the matched spacetimes are initially untrapped and
collapse, forming an apparent horizon and an asymptotically AdS black
hole with the corresponding topology.
\end{proposition}
\subsection{Spherical topology ($b=1$ and $k=0$)}
In this case, $f(R)=R$ and
the matching conditions become (for $p=\gamma\rho$ and $t(\tau)=\tau$):
\begin{eqnarray}
\label{equs}
 \dot{u}& \stackrel{\sigma}{=} &\varepsilon\frac{1 + \gamma}
 {\varepsilon\overline{\epsilon}\epsilon - \dot{a}R},
\\
\label{eqrs}
 r &\stackrel{\sigma}{=} &aR,\\
\label{eqms}
 m &\stackrel{\sigma}{=} &\frac{1}{2}a\dot{a}^2R^3 -\frac{\Lambda}{6}a^3R^3,\\
\label{eqRs}
 \dot{R}& \stackrel{\sigma}{=} &\varepsilon\overline{\epsilon}\epsilon\frac{\gamma}{a}.
\end{eqnarray}
Condition \eqref{spcond} must also be satisfied for the matching to
remain valid and, for spherical geometry, it can be written as
\begin{equation}\label{spconds}
\varepsilon \dot{a}R -  \overline{\epsilon}\epsilon < 0.
\end{equation}
In turn, the energy condition \eqref{wec}, for $\dot{u}>0$, gives
\begin{equation}\label{wecs}
 \varepsilon(\varepsilon\dot{a}R-\overline{\epsilon}\epsilon)\dot{m}\leq0.
\end{equation}
So, as long as condition \eqref{spconds} is satisfied, the $m$
function will grow for incoming radiation and will decrease for
outgoing radiation, as expected.

\begin{figure}[h!]
\includegraphics[scale=0.6]{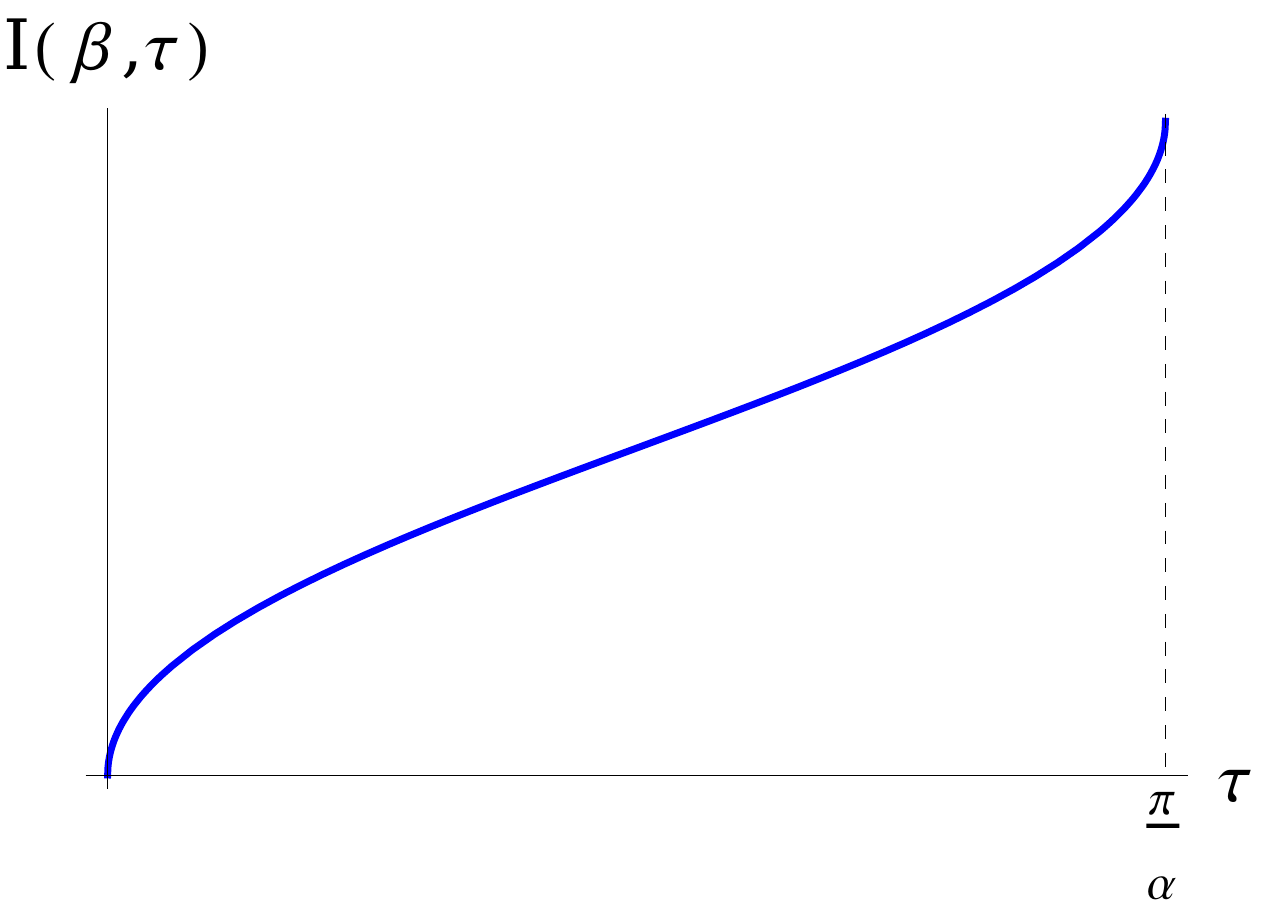}
\centering \caption{Plot of the function $I(\beta,\tau)$, for some
$\beta\in]1/3,2/3[$, i.e. some $\gamma\in]0,1[$.} \label{Itl0}
\end{figure}

\begin{figure}[h!]
\includegraphics[scale=0.6]{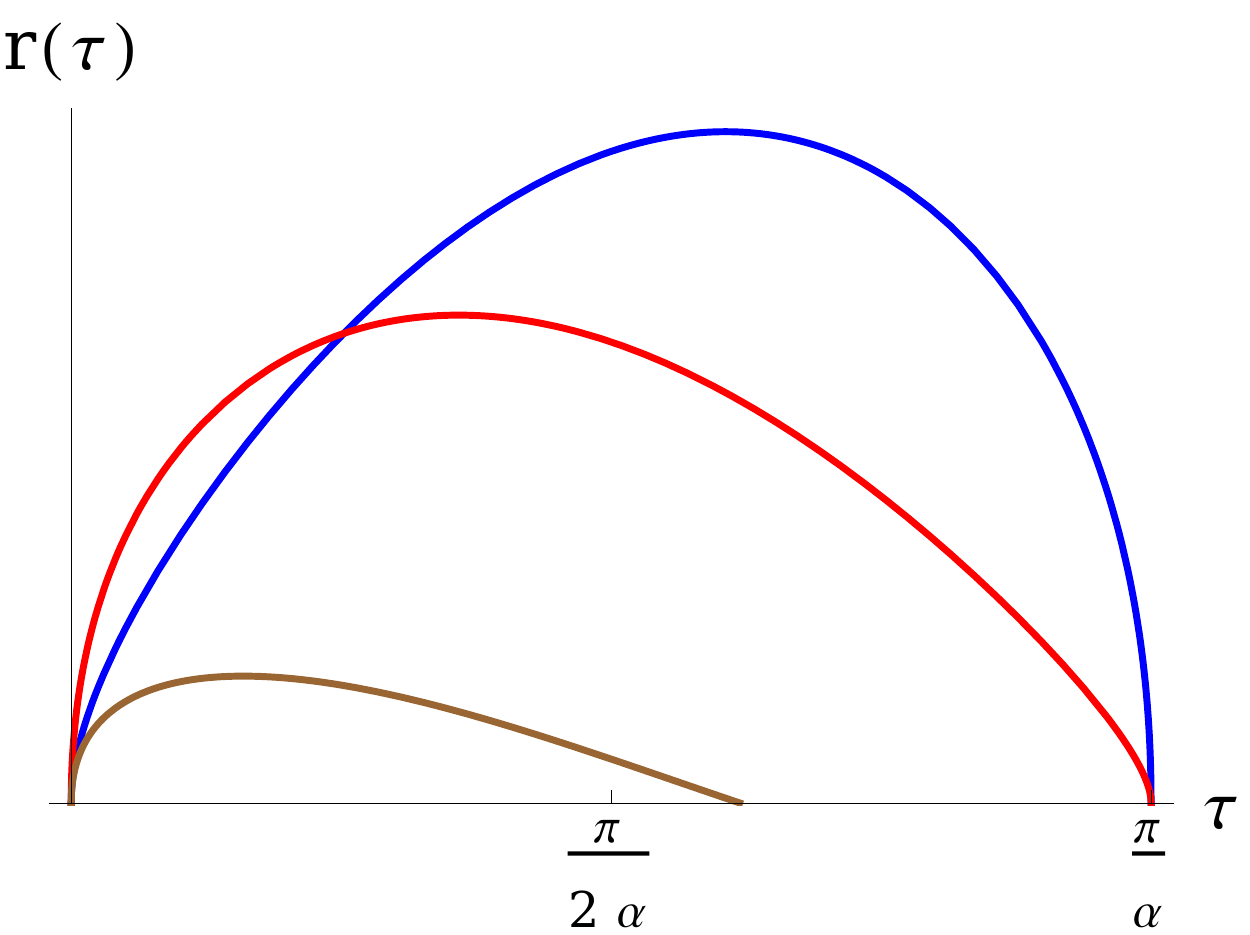}
\centering \caption{Examples of the function $r(\tau)$ for a fixed
$\gamma$ satisfying: (i)
$\varepsilon\overline{\epsilon}\epsilon\gamma>0$ (blue); (ii)
$\varepsilon\overline{\epsilon}\epsilon\gamma<0$ and some
$R_0>|\varepsilon\overline{\epsilon}\epsilon\gamma I(\beta,\tau)|$,
for all $\tau$ (red); (iii)
$\varepsilon\overline{\epsilon}\epsilon\gamma<0$ and
$R_0=|\varepsilon\overline{\epsilon}\epsilon\gamma
I(\beta,\tau_S)|$, for a finite $\tau_S$. Note that only in the last
case do we have collapse before $\tau=\pi/\alpha$.} \label{rtl0v}
\end{figure}

We now proceed with the study of the dynamics. Since $r=aR$, we need
to study both $a$ and $R$, in order to know if the system collapses
or expands. In \cite{FayosVilla}, given that $\Lambda=0$, it was
possible to choose a strictly increasing function $a$, so that $R$
was the function that decided if there was collapse or not. Here,
this is not the case. In fact, the solution of Friedmann's equation,
for $\Lambda<0$, can be written as
\begin{equation}
\label{scalefact}
 a(t) = a_0\sin^{\frac{2}{3(1+\gamma)}}(\alpha t),
\end{equation}
with $\alpha= (\gamma+1)\sqrt{3|\Lambda|}/2$ and $\alpha t
\in[0,\pi]$. This function increases up to a maximum $a_0$, that we
set as $a_0=1$ without loss of generality, and then decreases to
zero. We now need the solution for $R$, at the boundary, which is
given by
\begin{equation}
 R(\tau)  \stackrel{\sigma}{=} R_0 + \varepsilon\overline{\epsilon}\epsilon\gamma\int \frac{d\tau}{a(\tau)}
 = R_0 + \varepsilon\overline{\epsilon}\epsilon\gamma I(\beta,\tau),
\end{equation}
where $\beta = \frac{2}{3(1+\gamma)}$ and
\begin{equation}
\label{integralI}
 I(\beta,\tau) = \int \frac{d\tau}{\sin^{\beta}(\alpha\tau)}.
\end{equation}
For $\beta\geq1$, this integral does not converge for
$\alpha\tau\in[0,\pi]$, so we restrict $\gamma$ to the interval
$$-1/3<\gamma<1,$$
which means that $1/3<\beta<1$. Because of our freedom to choose
$R_0$, we can always set $I(\beta,0)=0$, by a convenient choice of
limits in the integral \eqref{integralI}, and then $R_0\ge 0$. With
this choice, we can then solve the integral giving
\begin{equation}
\begin{split}
 I(\beta,\tau)
 =\frac{\sqrt\pi\,}{2\alpha}\frac{\Gamma(1/2-\beta/2)}{\Gamma(1-\beta/2)}
 - \frac{\cos(\alpha\tau)}{\alpha}
 {}_2F_1\bigg(\frac{1}{2},\frac{1+\beta}{2},\frac{3}{2},\cos^2(\alpha\tau)\bigg),
\end{split}
\end{equation}
where $\Gamma$ is the gamma function and $_2F_1$ the hypergeometric
function. Although it may look messy, $I(\beta,\tau)$ is a well
behaved function for $\alpha\tau\in[0,\pi]$. Since
$\dot{I}(\beta,\tau)= \sin^{-\beta}(\alpha\tau)>0$, we know that $I$
grows until a maximum value at $\alpha\tau =\pi$ and an example is
shown in Figure \ref{Itl0}. The limit of $I$,
as $\tau$ approaches $\pi/\alpha$, is
\begin{equation}\label{lims}
 \lim_{\tau\to \pi/\alpha}I(\beta,\tau) = \frac{\sqrt{\pi}}{\alpha}\frac{\Gamma(1/2-\beta/2)}{\Gamma(1-\beta/2)}.
\end{equation}
When $\varepsilon\overline{\epsilon}\epsilon\gamma<0$, we want $R_0$
to be larger than this limit so that $R>0$ in the interval
$\alpha\tau\in[0,\pi]$. If, however, $R_0$ is not larger than this
limit, the collapse will happen before $\tau = \pi/\alpha$, at $\tau
= \tau_S$ defined by the equation
\begin{equation}\label{limsr}
R(\tau_S) = 0 \Rightarrow R_0 =
-\varepsilon\overline{\epsilon}\epsilon\gamma I(\beta,\tau_S).
\end{equation}
After fixing the value of $R_0$ of the initial system, this equation
can be solved numerically to obtain the collapse time $\tau_S$. For
a fixed $\beta$, equation \eqref{integralI} can be integrated to
obtain $R(\tau)$ which, together with \eqref{scalefact} and
\eqref{eqrs}, gives $r(\tau)$. Various examples of the evolution of
$r(\tau)$ can be seen in Figure \ref{rtl0v}.

Let us now examine how the dynamics is restricted by conditions
\eqref{spconds} and \eqref{wecs}. In order to do that, we need to
calculate $\dot{r}$ as
\begin{equation}\label{maxs}
 \dot{r} =\dot{a}R+a\dot{R} = \dot{a}R +\varepsilon\overline{\epsilon}\epsilon\gamma.
\end{equation}
According to equation \eqref{spconds}, $\dot{r}$ must satisfy
\begin{equation}\label{epsm1}
 \dot{r} < \overline{\epsilon}\epsilon(1+\gamma)
\end{equation}
for incoming radiation ($\varepsilon=1$) and
\begin{equation}\label{epsm2}
 \dot{r} > -\overline{\epsilon}\epsilon(1+\gamma)
\end{equation}
for outgoing radiation ($\varepsilon=-1$). From these results, it
can be seen that $\dot{r}  = \pm(1+\gamma)$ corresponds to a special
location which occurs for $\dot{a}R=\pm1$. To understand what is
happening, we need to discuss the marginally trapped surfaces of the
system which, as we know from Section 2, are defined in the exterior
by the equation $\chi=0$ and, in the interior, by $\dot{a}^2f^2-f'^2
= 0$. From the matching conditions, these equations are
equivalent at the matching surface, so that the marginally trapped
surfaces are continuous across the boundary, as expected. As was
seen in Section \ref{secext}, there will be an apparent horizon in
the exterior which is spacelike (i.e. a dynamical horizon) for
$b=1$, $\Lambda<0$ and $\varepsilon=1$.
The matching surface crosses the apparent horizon precisely when
\begin{equation}
\label{AH-eq}
 \dot{a}^2R^2 = 1,
\end{equation}
which has two solutions, satisfying \eqref{epsm1} and \eqref{epsm2},
for $\dot{r}=\pm(1\mp\gamma)$. In turn, these solutions limit the
possible $\tau$ values. In order to analyse this, we also need to
study the restriction to the $\gamma$ values which come from
condition \eqref{wecs}.

Before doing that, we note a difference between our approach and the
work of Fayos et al. \cite{FayosVilla}: While they have the initial
time at some $\tau>0$ and can have $R_0<0$, we fix the initial time
to be $\tau=0$ and have $R_0\ge 0$. So, we now analyse, separately,
the cases $R_0>0$ and $R_0=0$.

For $R_0 > 0$, we have that $\dot{r}$, from \eqref{maxs}, can be
written as
\begin{equation}
 \dot{r}(\tau) =R_0\sqrt \frac{|\Lambda|}{3}\sin^{\frac{-1-3\gamma}{3(1+\gamma)}}(\alpha\tau)\cos(\alpha\tau)+
 \varepsilon\overline{\epsilon}
 \epsilon\gamma\bigg[1+\sqrt\frac{|\Lambda|}{3}\sin^{\frac{-1-3\gamma}{3(1+\gamma)}}(\alpha\tau)\cos(\alpha\tau)I(\beta,\tau)\bigg].
\end{equation}
When $\tau\rightarrow0$, the first term diverges since we restricted
$\gamma\in~ ]-1/3,1[$, so $\dot{r}\rightarrow+\infty$. In that case,
this means that,  near $\tau=0$, we need $\varepsilon=-1$ in order
to satisfy condition \eqref{epsm2} for any
$\overline{\epsilon}\epsilon$. This solution must also satisfy the
energy condition \eqref{wecs} which tells us that, for
$\varepsilon=-1$, we need $\dot{m}\leq0$ for a timelike surface. The
$m$ function on $\sigma$ can be calculated from \eqref{eqms} as
\begin{equation}
\label{masseq}
 m(\tau) =R^3\frac{|\Lambda|}{6}\bigg[\sin^{\frac{-2\gamma}{1+\gamma}}(\alpha\tau)\cos^2(\alpha\tau)+\sin^{\frac{2}{1+\gamma}}
          (\alpha\tau)\bigg].
\end{equation}
The dominating term in this function, when we take the limit
$\tau\rightarrow0$, becomes
\begin{equation}
 m(\tau)\sim R_0^3\frac{|\Lambda|}{6}\sin^{\frac{-2\gamma}{1+\gamma}}(\alpha\tau)\cos^2(\alpha\tau).
\end{equation}
So, in order to have $\dot{m}\leq0$, near $\tau=0$, we need to have
$$\gamma>0.$$
Applying the same approach near $\tau=\pi/\alpha$, we get the same
restriction to the $\gamma$ values but with $\varepsilon=1$. This
garantees $\dot{m}\leq0$ in the whole allowed $\tau$ interval. This
is due to the fact that $m$, in \eqref{masseq}, has always a minimum
which occurs when \eqref{spconds} becomes an equality.

Unlike Fayos et al. in \cite{FayosVilla}, we can't exclude any
specific combination of $\varepsilon$ and
$\overline{\epsilon}\epsilon$ so, for $R_0>0$, we have the following
4 possible cases:
\begin{itemize}
\item \textbf{Case $\varepsilon=1$, $\overline{\epsilon}\epsilon=1$:}
In this case, $\dot{r}<1+\gamma$, which means that $\dot{r}$ can
start as positive, at some $\tau_i$, but the system will reach a
maximum radius before collapsing. The valid interval for $\tau$ is
$\tau\in]\tau_i,\pi/\alpha]$, where $\tau_S=\pi/\alpha$ is the time
of the singularity formation and $\tau_i$ is defined by
$\dot{a}(\tau_i)R(\tau_i) =1$. Since $\varepsilon=1$, then
$\dot{m}\ge 0$ and $m$ diverges at $\tau_S$.
The solutions of equation \eqref{AH-eq}, in this case, are $\dot{r}
= \pm1 + \gamma$, where $\dot{r} = 1+\gamma$ is the surface which
limits the $\tau$ values (as the matching is not valid beyond it)
and $\dot{r} =-1+\gamma$ corresponds to the apparent horizon which
happens at $\tau_H$ given by $\dot{a}(\tau_H)R(\tau_H) =-1$, after
the maximum of expansion at $\dot{r}=0$. This case, in particular,
is illustrated in Figure \ref{rtl0}.

\begin{figure}[h!]
\includegraphics[scale=0.6]{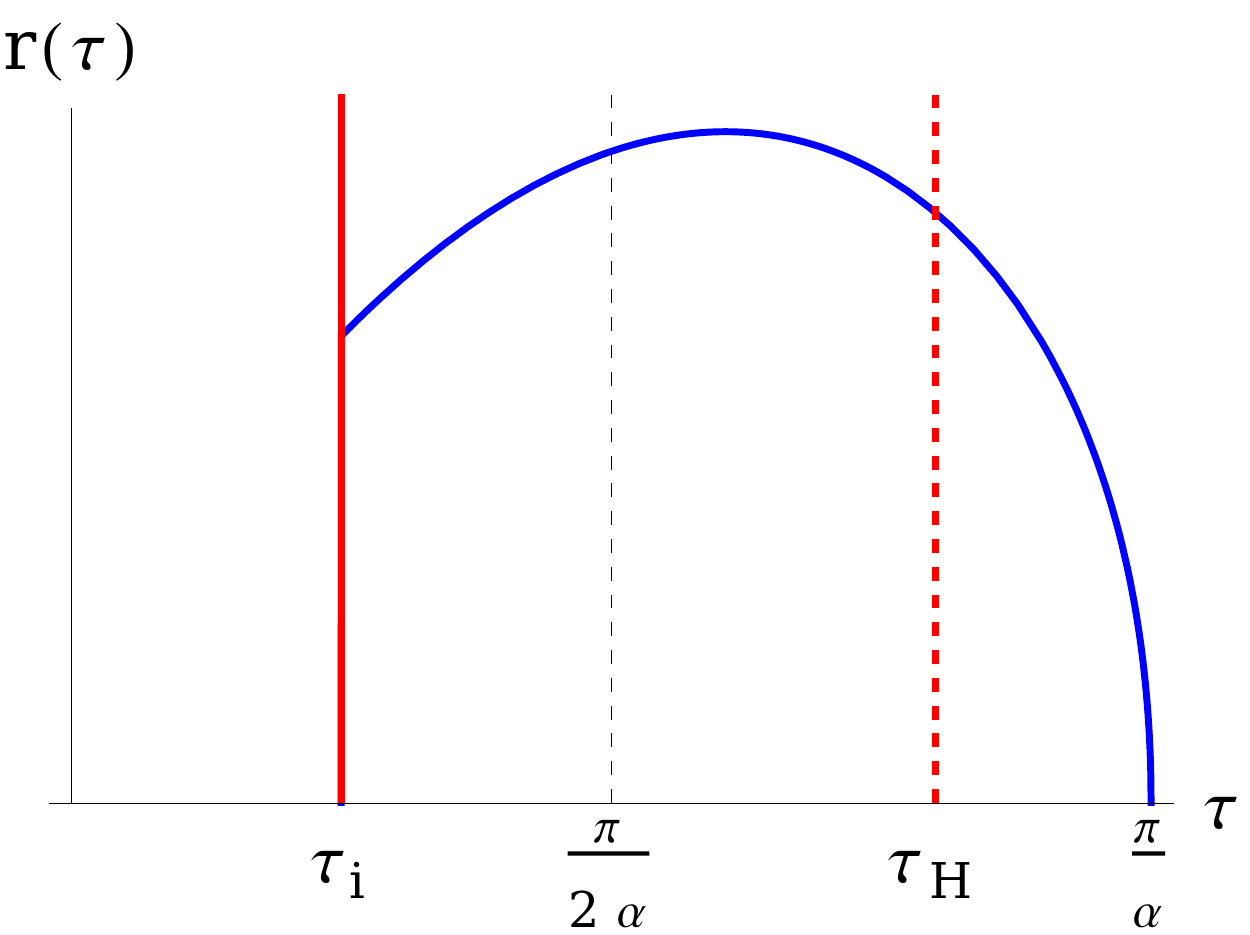}
\centering \caption{Example of the evolution of $r(\tau)$ for
$\varepsilon=\overline{\epsilon}\epsilon=1$. The red line (defined
at $\tau_i$) corresponds to the surface that limits the $\tau$
values and the red dashed line (defined by $\tau_H$) corresponds to
the time of the apparent horizon formation.} \label{rtl0}
\end{figure}

\item \textbf{Case $\varepsilon=1$, $\overline{\epsilon}\epsilon=-1$:}
In this case, $\dot{r}<-1-\gamma$, which means that $\dot{r}$ will
always be negative and the system will collapse. The valid interval
for $\tau$ is $\tau\in]\tau_i,\pi/\alpha]$, where $\tau_i$ is
defined by $\dot{a}(\tau_i)R(\tau_i)=-1$. Since $\varepsilon=1$, $m$
will again increase and diverge at $\tau_S=\pi/\alpha$.
The solutions of equation \eqref{AH-eq}, in this case, are given by
$\dot{r} = \pm1 - \gamma$, where $\dot{r}= 1-\gamma$ corresponds to
a surface that is outside the valid interval for $\tau$ and
$\dot{r}= -1-\gamma$ gives the surface that limits the $\tau$
values. Therefore, the system does not form an apparent horizon but
the collapse can, instead, be happening inside the trapped region.
Since $\varepsilon\overline{\epsilon}\epsilon=-1$, if $R_0$ is
smaller than the limit \eqref{lims}, the collapse occurs before
$\pi/\alpha$ at a time defined by equation \eqref{limsr}. We note
that this case was not included in \cite{FayosVilla}, since
condition \eqref{spconds} is not satisfied for $\Lambda=0$.

\item \textbf{Case $\varepsilon=-1$, $\overline{\epsilon}\epsilon=1$:}
Now $\dot{r} > -1-\gamma$, so $\dot{r}$ will start positive but will
end up negative. The system will expand up to a maximum radius
before reaching a null surface at $\tau_f$ defined by
$\dot{a}(\tau_f)R(\tau_f)=-1$. The valid interval for $\tau$ is
$[0,\tau_f[$, where $\tau=0$ corresponds to the starting
singularity. Since $\varepsilon=-1$, $m$ will decrease from
infinity, at $\tau=0$, to a finite value at $\tau_f$.
The solutions of equation \eqref{AH-eq}, in this case, are given by
$\dot{r} = \pm1-\gamma$, where $\dot{r}=-1-\gamma$ is the limiting
surface which arises at $\tau_f$. In turn, $\dot{r}=1-\gamma$
corresponds to a (past) marginally trapped surface, which occurs
before the time of maximum expansion at $\dot{r}=0$.
Again, since $\varepsilon\overline{\epsilon}\epsilon=-1$, if
$R_0<\gamma I(\beta,\tau_f)$, then the collapse occurs at $\tau_S$
defined by equation \eqref{limsr}. In this case, the mass $m$ goes
to zero at $\tau_S$ and we have a model of evaporation of a
white hole.

\item \textbf{Case $\varepsilon=-1$,
$\overline{\epsilon}\epsilon=-1$:} This time $\dot{r} >1+\gamma$, so
$\dot{r}$ will be always positive. The system will always expand
until it reaches a null surface at $\tau_f$, defined by
$\dot{a}(\tau_f)R(\tau_f)=1$. The valid interval for $\tau$ is
$[0,\tau_f[$, where $\tau=0$ corresponds again to the starting
singularity. Also, due to $\varepsilon=-1$, $m$ will decrease from
infinity, at $\tau=0$, to a finite value at $\tau_f$.
Equation \eqref{AH-eq} now gives $\dot{r} =\pm1 + \gamma$, where
$\dot{r} = 1+\gamma$ corresponds to a surface which limits the possible
$\tau$ values, and $\dot{r} =-1+\gamma$ is a surface that is outside
 the valid interval for $\tau$. Thus, the system expands from a
singularity but never crosses any horizon.
\end{itemize}
In all four cases above, except the second, the dynamics is
qualitatively similar to the corresponding $\Lambda=0$ cases
analized in \cite{FayosVilla}.

For $R_0=0$, however, we need a different reasoning. This time, $R$
can be written as $R(\tau) =
\varepsilon\overline{\epsilon}\epsilon\gamma I(\beta,\tau).$ Since
$I(\beta,\tau)$ is positive, we must have
$\varepsilon\overline{\epsilon}\epsilon\gamma>0$ so that $R>0$. In
this case, \eqref{eqms} gives
\begin{equation}
 m(\tau) =\varepsilon\overline{\epsilon}\epsilon\gamma \frac{|\Lambda|}{6}(I(\beta,\tau))^3\bigg[\sin^{\frac{-2\gamma}{1+\gamma}}
       (\alpha\tau)\cos^2(\alpha\tau)+\sin^{\frac{2}{1+\gamma}}(\alpha\tau)\bigg].\\
\end{equation}
This case has $m(0)=0$ (i.e. the collapse starts in AdS spacetime)
but, now, $m$ increases until it diverges at $\tau=\pi/\alpha$. So,
we get $\dot{m}>0$ which, by the energy condition \eqref{wecs},
means that we must always have $\varepsilon=1$. Therefore, we obtain
from \eqref{maxs}
\begin{equation}
 \dot{r}(\tau) = \varepsilon\overline{\epsilon}
 \epsilon\big[\gamma+\gamma I(\beta,\tau)\dot{a}\big] = \varepsilon\overline{\epsilon}\epsilon\bigg[\gamma+
 \gamma\sqrt\frac{|\Lambda|}{3}\sin^{\frac{-1-3\gamma}{3(1+\gamma)}}(\alpha\tau)\cos(\alpha\tau)I(\beta,\tau)\bigg],
\end{equation}
and, in order to satisfy \eqref{epsm1}, we need
$\overline{\epsilon}\epsilon\gamma I(\beta,\tau)\dot{a} <
\overline{\epsilon}\epsilon$. By estimating this inequality, we get
$\lim_{\tau\rightarrow0}\gamma I(\beta,\tau)\dot{a}<1/2$, for
$-1/3<\gamma<0$, which means that we need
$\overline{\epsilon}\epsilon=1$. This, together with the condition
$\varepsilon\overline{\epsilon}\epsilon\gamma>0$, implies that we
can have $R_0=0$ only when $\gamma>0$, $\varepsilon=1$ and
$\overline{\epsilon}\epsilon=1$. This corresponds to a system that
starts with $m\stackrel{\sigma}{=}0$, expands up to a maximum radius
and, then, recollapses to a singularity with a diverging $m$. The
apparent horizon, for this case, forms when $\dot{r} = -1+\gamma$.
\subsection{Toroidal topology ($b=0$ and $k=-1$)}
In this case, $f=e^{\pm R}$ and the matching conditions (with
$t(\tau)=\tau$ and $p=\gamma\rho$) reduce to
\begin{eqnarray}
\label{equm1}
 \dot{u} &\stackrel{\sigma}{=}&\varepsilon\frac{1 + \gamma}
 {\pm\varepsilon\overline{\epsilon}\epsilon - \dot{a}}e^{\mp R},\\
\label{eqrm1}
 r& \stackrel{\sigma}{=}& ae^{\pm R},\\
\label{eqmm1}
 m &\stackrel{\sigma}{=} &\frac{1}{2}ae^{\pm3R}(\dot{a}^2-1-\frac{\Lambda}{3}a^2),\\
\label{eqRm1}
 \dot{R} &\stackrel{\sigma}{=} &\varepsilon\overline{\epsilon}\epsilon\frac{\gamma}{a}.
\end{eqnarray}
Condition \eqref{spcond} is now
\begin{equation}\label{spcond2}
 \pm\overline{\epsilon}\epsilon - \varepsilon \dot{a} > 0
\end{equation}
and the energy condition \eqref{wec}, on the matching surface, can
 be rewritten as
\begin{equation}\label{wecm1}
 \varepsilon(\pm\overline{\epsilon}\epsilon - \varepsilon \dot{a})\dot{m} \geq 0.
\end{equation}
For $k=-1$ and $\Lambda<0$, equation \eqref{eqa} does not admit, in
general, an explicit solution for open sets of $\gamma$, unlike the
case $k=0$. So, we will consider particular values of $\gamma\ne 0$
satisfying the dominant energy conditions, and for which explicit
expressions of the scale factor $a$ are known \cite{Harrison},
namely $\gamma=1/3$ and $\gamma=-2/3$. Furthermore, to simplify the
discussion we will focus on the collapsing dynamics of the spacetime
without a previous expansion phase.

As far as the exterior is concerned, we know already from
\cite{VillaVaydia} that the collapse of spacetimes \eqref{exterior} with toroidal topology can lead to black hole formation (for
$\varepsilon=1$). The difference here is that, by finding interiors to
such spacetimes, we can interpret those interiors as being the
sources of mass and radiation for the exterior. An interesting
consequence of this is that by imposing $\rho>0$ in the interior, by
\eqref{masscontinuity}, we must have $m>0$ at the boundary and,
locally, in the exterior. While, in \cite{VillaVaydia}, there could
be black hole formation with $m<0$ and still satisfying the dominant
energy conditions.

We also recall that, from \cite{VillaVaydia}, and as revised in
Section 2.2, the surfaces of transitivity of the exterior are
trapped in the region $2m(u)>-\Lambda r^3/3>0$ and the marginally
trapped tubes (when they exist) are non-timelike, having null parts
if $dm/du=0$ and spacelike parts if $dm/du>0$. This will help us to
find viable interiors next.
\subsubsection{Case ${\displaystyle \gamma=1/3}$}
In this case, the Friedmann equations admits the following
collapsing explicit solution:
\begin{equation}
\label{aevol}
 a(t) = \frac{\sqrt{2}}{\alpha}\sqrt{1-\cos(\alpha t+\theta_0)+\beta\sin(\alpha
 t+\theta_0)},
\end{equation}
where $\alpha = 2\sqrt{|\Lambda|/3}$ and $\beta$, $\theta_0$ are
constants defined by the initial conditions. In turn, from
\eqref{eqRm1}, we get
\begin{equation}
\label{needthis}
 R(\tau) = R_0 + \frac{1}{3}\varepsilon\epsilon\overline{\epsilon}\int_0^\tau \frac{dt}{a(t)} :=
 R_0 + \frac{1}{3}\varepsilon\epsilon\overline{\epsilon}L(\tau).
\end{equation}
The function $L(\tau)$ can be obtained in terms of products
involving trigonometric functions and Appell hypergeometric
functions. Choosing appropriate intervals of continuity for
$L(\tau)$ where $\dot a(0)=0$ and $\dot a(\tau)<0$, for $\tau>0$, we
then fix $\beta=1$ and $\theta_0=3\pi/4$ so that
$\alpha\tau\in[0,3\pi/4]$.
In that case, the function $L(\tau)$ can be written in the
considerably simpler form
\begin{equation}
 L(\tau) = \sqrt{2\sqrt{2}-2}\, F\bigg(\frac{\alpha\tau}{2}, 4 - 2
 \sqrt{2}\bigg),
\end{equation}
where $F$ is an elliptic function of the first kind. This function
is finite and continuous for $\alpha\tau\in[0,3\pi/4]$. Now that we
have a well-behaved $R$ function we can just ignore it and focus
only on $a(\tau)$. This is because $r=af=ae^{\pm R}$ and so, as long
as $R$ is finite, the collapse is decided solely by the function
$a(\tau)$. Furthermore, the apparent horizon is defined by the
equation $f'^2-\dot{a}^2f^2=0$ which, for this case, reduces to
$\dot{a}^2 = 1$ so, again, we just need $a(\tau)$. The only
condition where $f$ is necessary is \eqref{spcond2}, where we need
to know which $f$ function should be chosen.

Now, from \eqref{aevol}, we get
\begin{equation}
 \dot{a}(\tau) = \frac{\cos{(\alpha\tau+\pi/4)}-\sin{(\alpha\tau+\pi/4)}}{\sqrt{2}\sqrt{1+\cos{(\alpha\tau+\pi/4)}+\sin{(\alpha\tau+\pi/4)}}}.
\end{equation}
This function starts at zero and decreases to minus infinity as it
approaches $\alpha\tau=3\pi/4$. The apparent horizon occurs
for $\alpha\tau = \pi/2$. In turn,
\begin{equation}
 m(\tau) = \frac{1}{2}af^3\bigg(\dot{a}^2-1 +\frac{\Lambda a^2}{3}\bigg) =
 \frac{1}{2\alpha}\frac{e^{\pm3R(\tau)}}{\sqrt{2+2\sqrt{2}\cos{(\alpha\tau)}}}.
\end{equation}
The function between the brackets starts as positive and it
is ever increasing, diverging at $\alpha\tau=3\pi/4$. To know
$\dot{m}$ we need to know if $f$ increases or decreases, which
depends purely on the sign of
$\varepsilon\overline{\epsilon}\epsilon$. When
$\varepsilon\overline{\epsilon}\epsilon$ is positive (resp. negative), we have that
$f$ increases (resp. decreases) for $f=e^R$. 
Furthermore, $f$
only dominates the dynamics near $\tau=0$ being overpowered near
$\alpha\tau=3\pi/4$, where $m$ diverges. Then, we focus on the cases
where $f$ increases, as they are the only cases that cause collapse
and always satisfy condition \eqref{wecm1}, for a fixed
$\varepsilon$ (which, in this case, is $\varepsilon=1$, just like the cases of black hole formation of \cite{VillaVaydia}).

So, now that we have $\dot{m}>0$, $\dot{f}>0$ and $\varepsilon=1$,
the condition \eqref{spcond2} implies
\begin{equation}\label{spcondtt}
 \pm\overline{\epsilon}\epsilon - \dot{a} >0.
\end{equation}
This implies that we have two possible cases allowing for
$\dot{f}>0$, namely $f=e^{\pm R},\overline{\epsilon}\epsilon=\pm 1$.
We then get that condition \eqref{spcondtt} becomes $1-\dot{a}>0$
which is always satisfied for both cases, given that $\dot{a}<0$.
So, the possible cases are
$f=e^{\pm R}$ for $\overline{\epsilon}\epsilon=\pm1$ and the matching
is valid for $\alpha\tau\in[0,3\pi/4]$,
with the apparent horizon being formed at $\alpha\tau_{H}=\pi/2$. As
was specified in Section \ref{secext}, this apparent horizon is
always spacelike, so it is an example of a dynamical horizon.
%
\subsubsection{Case ${\displaystyle \gamma=-2/3}$}
%
In this case, the Friedmann equations admit the explicit solution
\begin{equation}
\label{afactor}
 a(t) = \frac{\beta}{2\alpha^2}[1-\cos{(\alpha t + \theta_0)}]+\frac{1}{\alpha}\sin{(\alpha t
 +\theta_0)},
\end{equation}
where $\alpha = \sqrt{|\Lambda|/3}$ and $\beta$, $\theta_0$ are again defined by the initial conditions.
This function starts increasing
from zero at $\alpha t + \theta_0 =0$, reaches a maximum at $\alpha t + \theta_0 = \pi - \arctan(2)$ and then decreases to zero at
$\alpha t + \theta_0 = 2\pi-2\arctan(2)$.
As in the previous section, we need to control
\begin{equation}
 R(\tau) = R_0 -\frac{2}{3}\varepsilon\overline{\epsilon}\epsilon\int_0^\tau\frac{dt}{a(t)} :=
 R_0 -\frac{2}{3}\varepsilon\overline{\epsilon}\epsilon K(\tau)
\end{equation}
and, by similar reasons as in the previous case, we now choose
$\beta=\alpha$ and $\theta_0=\pi-\arctan2$ so that
$\alpha\tau\in[0,\pi-\arctan2]$. With this choice, we get
\begin{equation}
 K(\tau) = \frac{1}{2}\ln
 5-\ln\bigg[1+2\tan\bigg(\frac{-\alpha\tau+\arctan2}{2}\bigg)\bigg],
\end{equation}
which is positive and always increasing for
$\alpha\tau\in[0,\pi-\arctan 2[$. However, this function diverges at
$\alpha\tau=\pi-\arctan2$, which causes $R$ to be negative if we
choose $\varepsilon\overline{\epsilon}\epsilon=1$. So, for the
collapse to be possible with positive $R$, we need to choose
$\varepsilon\overline{\epsilon}\epsilon=-1$. The function $r=af$
still goes to zero at $\alpha\tau=\pi-\arctan2$, even if we choose
the diverging function $f=e^R$.

Now, from \eqref{afactor}, we get
\begin{equation}
 \dot{a}(\tau) = -\frac{1}{2}\sin(\alpha \tau-\arctan2) -\cos(\alpha
 \tau-\arctan2),
\end{equation}
which, unlike the case of the previous section, does not diverge but
instead starts at zero, reaches a minimum at $\alpha\tau = \arctan 2
+\arctan\frac{1}{2}$ and then increases again. Because of this, the
marginally trapped surface equation $\dot{a}^2=1$ has two solutions,
$\alpha\tau=\arctan2$ and $\alpha\tau=\pi-\arctan2$, where the
latter corresponds to the collapse time.

In turn, the mass function is
\begin{equation}
 m(\tau) = \frac{1}{2}af^3\bigg(\dot{a}^2-1+\alpha^2 a^2\bigg) = \frac{1}{16\alpha}e^{\pm3R(\tau)}\big(7+4\sqrt{5}\cos{(\alpha \tau)}+5\cos{(2\alpha
 \tau)}\big),
\end{equation}
which, given $\alpha\tau\in[0,\pi-\arctan2]$, always decreases to
zero for $f=e^{-R}$, and increases to a maximum before decreasing to
zero for $f=e^R$. Therefore, we will choose $f=e^{-R}$ which is the
only solution allowing for a fixed $\varepsilon$ in the whole $\tau$
interval, meaning $\varepsilon=-1$, according to the energy
condition \eqref{wecm1}. We then choose
$\varepsilon\overline{\epsilon}\epsilon=-1$, i.e.
$\overline{\epsilon}\epsilon=1$.

With $f=e^{-R}$, condition \eqref{spcond2} becomes
\begin{equation}
 1-\dot{a} >0,
\end{equation}
which is always satisfied, since $\dot{a}<0$. Therefore, we have a
collapse in the whole interval $\alpha\tau\in[0,\pi-\arctan2]$ and,
as $\varepsilon=-1$, we have a marginally past trapped surface that
forms at $\alpha\tau=\arctan2$. So, this case corresponds to a white
hole.
%
\subsection{Higher genus topology ($b=-1$ and $k=-1$)}
%
As far as we know, this is the first time that the 
spacetimes \eqref{exterior} with $b=-1$ are considered  in the context of collapse. So,
even the study of particular cases seem to be of interest.
As we now have $f=\cosh R$, the matching conditions become (for
$t(\tau)=\tau$ and $p=\gamma\rho$):
\begin{eqnarray}
\label{equhm1}
 \dot{u} &\stackrel{\sigma}{=}& \varepsilon\frac{1 + \gamma}
 {\varepsilon\overline{\epsilon}\epsilon\sinh R - \dot{a}\cosh R},\\
\label{eqrhm1}
 r &\stackrel{\sigma}{=} & a\cosh R,\\
\label{eqhm1}
 m &\stackrel{\sigma}{=}& \frac{1}{2}a\cosh^3 R(\dot{a}^2 - 1
 -\frac{\Lambda}{3}a^2),\\
\label{eqRhm1}
 \dot{R} &\stackrel{\sigma}{=} &\varepsilon\overline{\epsilon}\epsilon\frac{\gamma}{a}.
\end{eqnarray}
Condition \eqref{spcond} becomes
\begin{equation}\label{spcondhm1}
 \overline{\epsilon}\epsilon\tanh R - \varepsilon \dot{a} > 0
\end{equation}
and the dominant energy condition is
\begin{equation}\label{wechm1}
 \varepsilon(\overline{\epsilon}\epsilon\tanh R - \varepsilon \dot{a})m\geq
 0.
\end{equation}
As in the previous section, we will consider particular cases of
$\gamma$, namely $\gamma=1/3, \gamma=-1/3$ and $\gamma=-2/3$,
satisfying the dominant energy conditions, and focus on the
collapsing part of the spacetime dynamics. We know, from Section
2.2, that in this case, the surfaces of transitivity of the exterior
are trapped in the region $2m(u)+\Lambda r^3/3+r>0$ and the
marginally trapped tubes (when they exist) are timelike if
$1>|\Lambda|r^2$, spacelike if $1<|\Lambda|r^2$ (dynamical horizons)
and null if $dm/du=0$.
%
\subsubsection{Case $\gamma=1/3$}
%
In this case, the solutions for $a$ and $R$ are exactly the same as
in Section 4.2.1 and given by equations \eqref{aevol} and
\eqref{needthis}. As before, we just consider
$\alpha\tau\in[0,3\pi/4]$, with $\theta_0=3\pi/4$ and $\beta=1$.
In turn, the time $\tau_H$ of the apparent horizon formation is
given by
\begin{equation}
\label{hor-hyp}
 \chi = \sinh^2R - \dot{a}^2\cosh^2R = 0 \Rightarrow \tanh^2R
 =\dot{a}^2,
\end{equation}
which now depends on $R_0$ and on the sign of
$\varepsilon\overline{\epsilon}\epsilon$. Since $\dot{a}^2$ starts
at zero and diverges for $\alpha\tau=3\pi/4$ and $\tanh^2 R<1$, then
equation \eqref{hor-hyp} will always have a solution which is finite
if the derivative of $\tanh^2R$, at $\tau=0$, is larger than the
derivative of $\dot{a}^2$ at $\tau=0$.

Now, for the mass function, we get
\begin{equation}
 m(\tau) = \frac{1}{2\alpha}\frac{\cosh^3R(\tau)}{\sqrt{2+2\sqrt{2}\cos(\alpha\tau)}}.
\end{equation}
This time, $\dot{f}$ is positive for
$\varepsilon\overline{\epsilon}\epsilon=1$ and negative for
$\varepsilon\overline{\epsilon}\epsilon=-1$ (considering $R\geq0$).
Using the same argument as in the last section, since $m$ diverges
 at $\alpha\tau=3\pi/4$, we will consider
$\varepsilon=1$ which gives $\dot{m}>0$ and satisfies the energy
condition \eqref{wechm1} in the whole time interval. This fact,
together with condition \eqref{spcondhm1}, gives
$\varepsilon\overline{\epsilon}\epsilon=1$ and  tells us that we
have just one possible case: $\overline{\epsilon}\epsilon=1$.
In this case, \eqref{spcondhm1} becomes $\tanh R > \dot{a}$ which,
as long as $R$ is positive, is always satisfied in the considered
time interval. Then, the possible $\tau$ values are
$\tau\in[0,3\pi/4\alpha]$ and the apparent horizon is formed at a
time $\tau_H$ defined by the equation $\tanh R(\tau_H) =
-\dot{a}(\tau_H)$.
As $m$ diverges, we will always end up having a large enough horizon
radius $r_H$ where both $\chi=0$ and $|\Lambda|r_H^2>1$ are
verified. This means that we will always end up having a dynamical
horizon. As far as we are aware, this is the first example of the
formation of a dynamical horizon with higher genus topology.
%
\subsubsection{Cases $\gamma=-1/3$ and $\gamma=-2/3$ }
%
Using a similar procedure as in Section 4.2.2, we could not find
feasible models of collapse, in these cases. This is due to the fact
that condition \eqref{spcondhm1} did not remain valid after the
trapped surface formation and, therefore, the matching could not be
properly described. More details about these cases can be found in
\cite{Oliveira}.
%
\section{Conclusions}
In this paper, we have derived the matching conditions between the generalised Vaidya metric \eqref{exterior} and the FLRW metrics \eqref{interior} for perfect fluid source fields, across timelike hypersurfaces. We have then imposed a linear equation of state in the FLRW fluid and proved existence results for the matching, satisfying the weak and dominant energy conditions. In particular, we have constructed models of radiative gravitational collapse which result in the formation of asymptotically AdS black holes with spherical, toroidal and higher genus topologies. We have also found cases where, instead, a white hole forms. A summary of our results can be found in Table \ref{table-conc}. 
\begin{center}
	\begin{table}[h]
		\centering
		\resizebox{0.85\textwidth}{!}{%
		\begin{tabular}{ | M{2.5cm} | M{4.7cm} | M{5.2cm} | M{3.7cm} |}
			\cline{2-4}
			\multicolumn{1}{c|}{} & Black Hole Formation & White Hole Formation & Dynamical Horizon \\ \hline
			Spherical  & $\varepsilon=1,~\overline{\epsilon}\epsilon=\pm1,~0<\gamma<1$  & $\varepsilon=-1,~\overline{\epsilon}\epsilon=\pm1,~0<\gamma<1$           &  $\varepsilon=1,~0<\gamma<1$ \\ \hline
			Toroidal  &  $\varepsilon=1,~\overline{\epsilon}\epsilon=\pm1,~\gamma=1/3$   &  $\varepsilon=-1,~\overline{\epsilon}\epsilon=1,~\gamma=-2/3$               & $\varepsilon=1,~\gamma=1/3$ \\ \hline
			Higher Genus  & $\varepsilon=1,~\overline{\epsilon}\epsilon=\pm1,~\gamma=1/3$     & None found  & $\varepsilon=1,~\gamma=1/3$ when $r_H > 1/|\Lambda|$\\
			\hline
		\end{tabular}%
	}
		\caption{Table summarizing our results as well as the restrictions on the free parameters. The parameter $\varepsilon$ refers to the advanced or retarded character of the $u$ coordinate in \eqref{exterior}. The constants $\overline{\epsilon}$ and $\epsilon$ satisfy $\overline{\epsilon}\epsilon=\pm1$ by definition, see \eqref{normals}, and $\gamma$ is the parameter of the FLRW linear equation of state $p=\gamma \rho$.}\label{table-conc}
	\end{table}%
	\vspace{-10pt}
\end{center}

From the physical point of view, the spherically symmetric case of Section 4.1 contains a specially interesting subcase, given by $\varepsilon=1$, which models a compact region containing a fluid, with any $1>\gamma>0$, and having a radiating exterior. The fluid then collapses into a black hole while it exchanges energy with the exterior region.

Another interesting property that has been found is the existence of dynamical horizons in these settings. In particular, as far as we know, we provide the first example of the formation of a dynamical horizon with higher genus topology. A discussion about the importance and applications of dynamical horizons is given in \cite{Ashtekar} and includes different aspects of black hole mechanics, numerical relativity and mathematical physics beyond the Einstein-Maxwell theory.

Our work can be seen as an extension of the following past works: (i) \cite{Smith-Mann}, which considers the interior FLRW metric \eqref{interior} with $k=-1$ (in particular, the cases (b) and (c) of Table 1), having a dust source field, matched to an exterior metric \eqref{exterior} with $m(u)=$ const., $\Lambda<0$ and $b=-1$, i.e. to a Kottler metric with hyperbolic symmetry; (ii) \cite{MenaCollapse}, which takes the remaining non-spherical cases of Table 1, again with a dust source, and exteriors given by \eqref{exterior} with $m(u)=$ const., $\Lambda<0$ and $b=0,-1$, i.e. by Kottler metrics with planar or hyperbolic symmetry; (iii) \cite{FayosVilla}, which considers  perfect fluid FLRW metrics \eqref{interior}, with a linear equation of state and $k=1$, matched to the metrics \eqref{exterior} with $b=1$ and $\Lambda=0$, i.e. to  (radiating) Vaidya exteriors in spherical symmetry only.  

As a final remark, we note the matching conditions are local conditions and one can swap the role of the interior and exterior metrics. In that case, our generalised Vaidya spacetime could be seen as a compact region embedded in an evolving FLRW cosmological model. This setting would require a new analysis of the equations, but it could be potentially interesting to study e.g. the formation of primordial black holes in the early universe.
\section*{Acknowledgments}
We thank Jos\'e Senovilla for useful comments on the manuscript. This research was supported by Portuguese Funds through FCT -
Funda\c{c}\~ao para a Ci\^encia e a Tecnologia, within the projects UID/MAT/00013/2013 and PTDC/MAT-ANA/1275/2014 as well as the Ph.D. grant PD/BD/128184/2016. 

\end{document}